\documentclass[ ]{aastex701}
\usepackage{xcolor}
\usepackage{subcaption}
\usepackage{amsmath}
\usepackage{float}
\begin{document}
\def\be{\begin{equation}}
\def\ee{\end{equation}}
\def\ba{\begin{eqnarray}}
\def\ea{\end{eqnarray}}
\def\parderiv#1#2{\frac{\partial #1}{\partial #2}}

\def\rvec{{\bf r}}
\def\rvece{{\bf r_e}}
\def\rvecm{{\bf r_m}}
\def\omegavec{\boldsymbol{\omega}}
\def\Vvec{{\bf V}}
\def\Avec{{\bf A}}
\def\Rvec{{\bf R}}
\def\Dvec{{\bf D}}
\def\vvec{{\bf v}}
\title{A comparative study of time on Mars with lunar and terrestrial clocks}

\author[orcid=0000-0003-0719-3979,sname='North America']{Neil Ashby}  
\affil{National Institute of Standards and Technology, Boulder, CO 80305}
\email[show]{Neil.Ashby@Colorado.Edu}

\author[orcid=0000-0003-0575-7080,sname='North America']{Bijunath R. Patla}
\affil{National Institute of Standards and Technology, Boulder, CO 80305}
\email[show]{bijunath.patla@nist.gov}

\date{\today}

\begin{abstract}

As space exploration extends into cislunar space and further towards Mars, understanding the relativistic effects on clocks on Mars—particularly in relation to multi-body gravitational influences—becomes increasingly important for accurate clock synchronization.
This study estimates clock rates on Mars and compares them to those on the Moon and Earth. We find that, on average, clocks on Mars tick faster than those on the Earth's geoid by 477~$\mu\rm{s}~\rm{day}^{-1}$, with a variation of  226~$\mu\rm{s}~\rm{day}^{-1}$ over a Martian year. Additionally, there is an amplitude modulation of approximately
40~$\mu\rm{s}~\rm{day}^{-1}$ over seven synodic cycles. We also introduce a formalism that includes the effects of solar tides on the Earth-Moon system for predicting clock rates on the Moon and Mars more accurately when compared to using only Keplerian orbit approximations.
Our analysis quantifies the relativistic proper-time offsets among Martian, lunar, and terrestrial clocks, highlighting important implications for mission planning and the implementation of timekeeping systems on Mars.

\end{abstract}

\section{Introduction}
Mars resides in a gravitational environment fundamentally distinct from Earth's, characterized by its greater heliocentric distance, higher orbital eccentricity, and lower surface gravity. At an average distance of approximately 1.52 astronomical units ($\mathrm{AU}$) from the Sun, compared to Earth's $1\,\mathrm{AU}$, Mars experiences a weaker solar gravitational potential, which directly affects proper time rates through general relativistic time dilation. Additionally, Mars’s orbital eccentricity~(0.093) is significantly larger than Earth’s~(0.017), leading to substantial variations in heliocentric distance and gravitational potential over a Martian year. Its surface gravity is also five times weaker than that on Earth’s geoid.

These combined factors result in relativistic clock rate differences that are larger in magnitude and more variable than those observed in the Earth–Moon system. This makes Mars an invaluable natural laboratory for testing relativistic timekeeping models, validating gravitational time dilation predictions over varying orbital configurations, and developing autonomous interplanetary time synchronization frameworks. As human and robotic missions extend further into the solar system, Mars is a critical test bed for building scalable timekeeping infrastructure beyond the cislunar regime.

Space-faring nations,  as well as several commercial entities, are already advancing plans for crewed and resource-focused missions to the Moon. NASA and its international partners, hoping to leverage the experience and lessons learned from sustained lunar surface operations, are also developing a roadmap and requisite elements for future human missions to Mars and deep space exploration with the Moon to Mars architecture~\cite{artemis_20,moon_to_mars}.
Establishing a dedicated time standard for the Moon---analogous to Coordinated Universal Time (UTC) on Earth---would significantly support long-term lunar exploration, depending on the scale and scope of the planned activities involving communications, positioning, and navigation (CPNT) infrastructure in the future~\cite{ashbypatla24,kopeikin2024lunar,iau24,Turyshev_2025}. The framework proposed in~\cite{ashbypatla24} presents a scalable approach that can be adapted for Mars, which is the central focus of this paper. A comparative analysis of clock rates on the Moon and Mars in relation to Earth is essential to evaluate the scientific merits and address the technical challenges of expanding human activities beyond Earth. This analysis provides valuable insights into how different time scales can be related and synchronized with Coordinated Universal Time (UTC)~\cite{bipm_utc}.

In \cite{ashbypatla24}, the relativistic framework enabled comparison of clock rates on the Moon and cislunar Lagrange points with respect to clocks on Earth by using metric tensors appropriate for a locally freely falling frame. In this paper, we show how an independent  reference time can be established on Mars and, using available Mars gravity field data, derive the constant $L_M$, the rate offset of clocks on  Mars's areoid  with respect to the center of mass of Mars, corresponding to the constant $L_G$ defined for the Earth.  
%This requires the introduction of a locally inertial, freely falling reference frame whose origin is the center of mass of Mars. 
Then, $-L_M$ would define the linear rate of clocks on the  equipotential surface of Mars with respect to the clock at the  center.  For rate comparisons, it is assumed that the initial offsets between remote clocks can be determined with high accuracy through clock transport or the exchange of light signals.

Since observers moving with a clock can directly measure only proper time, estimating clock rates on Mars relative to those on Earth and the Moon requires comparing proper time intervals experienced at each location. To maintain generality and focus on the more dominant effects, we constrain our analysis to an accuracy threshold of 50~ns---corresponding to a  fractional frequency of approximately $5\times10^{-13}$, averaged over a day. Terms with smaller magnitudes are neglected, as are some periodic contributions, which are beyond the scope of the current estimates. We note that the fully relativistic coordinate transformations adopted by resolutions of the International Astronomical Union~(IAU) for atomic clock comparisons are accurate to at least $10^{-16}$ or better in fractional frequencies~\cite{iau03}.

In $\S2$ we provide a best estimate of $L_M$, the equivalent of $L_G$ for the Earth, using two different models.
We start $\S3$ with intuitive arguments to roughly estimate the proper time rate offset between Mars and Earth, which can amount to several hundred microseconds per day, due primarily to the two planets' differing distances from the sun, followed by formal treatment by choosing a metric that is appropriate for  the solar system barycentric celestial reference system (BCRS)~\cite{iau03}. 
$\S~4$  presents the tidal perturbations on the Earth-Moon system and a numerical integration scheme that can better match the predictions of rate offsets compared to the planetary and lunar ephemerides DE440
 ~\cite{park_2021}. Discussions and conclusions are presented in $\S~5$. 
The metric signature used is $(-1,1,1,1)$, with Greek indices running from 0 to 3, and a negative sign is assigned for gravitational potential. All the constants used are given in Table~\ref{tab:const}.
 
 %\footnote{here is a new footnote}

%-------------------
\section{ A local inertial frame for Mars and the constant $L_M$}

Construction of a locally inertial, freely falling frame  with a coordinate time defined by an ideal clock at the origin has been discussed in~\cite{fermi1922,ashbybertotti86,ashbypatla24}.  A clock that will realize the  International System of Units (SI) second is considered an ideal clock~\cite{bipm2019si}. 
We review here the development of the constant $L_G$ for the Earth, which is essential in defining coordinate time on the Earth.  The gravitational potential on Earth's geoid, including the monopole and quadrupole term, is denoted by $\sum\Phi_e$.  Viewed from an  Earth-centered inertial frame, the second-order Doppler shift on the equator due to Earth's rotation can be included, and the combined sum of these two effects may be 
%
%\be
%\left(\frac{\delta f}{f}\right)_{\rm Dop}=-\frac{\omega_e^2 a_e^2}{2 c^2},
%\ee
%
%where $\omega_e$ is Earth's angular rotation rate and $a_e$ is the equatorial radius.  
%
%Combining these two effects, their sum is 
expressed in terms of a constant $L_G$ or the effective potential $\Phi_0$~\cite{iau,iau03,petit2010}
\be
\label{eq:lg_def1} 
-L_G=\frac{\Phi_0}{c^2}=\frac{ \sum \Phi_e}{c^2}-\frac{\omega_e^2 a_e^2}{2 c^2}=-6.969290134 \times 10^{-10},
\ee
where $\omega_e$ is Earth's angular rotation rate, $c$ is the speed of light in vacuum, and $a_e$ is the equatorial radius  of the Earth.
$-L_G  = (dTT - dTCG)/dTCG $ in Equation~(\ref{eq:lg_def1}) is the fractional rate difference between a clock fixed on the geoid that realizes Terrestrial Time~(TT) on the rotating Earth, and a clock at the origin of the Geocentric Coordinate Reference System~(GCRS) that defines Geocentric Coordinate Time~(TCG) as per the resolutions adopted by the IAU~\cite{iau,iau03,petit2010}.  $TCG$ advances faster than $TT$ by a linear rate $L_G \sim~60.2~\mu$s ~day$^{-1}$, where a day is defined to be 86,400 SI seconds.

 This construction can be extended to define a local coordinate time in the vicinity of the Martian surface, see for example, ~\cite{BrumbergKopeikin1990}. Mars’s equivalent of the geoid—the areoid—is nearly a surface of effective hydrostatic equilibrium, and its equatorial radius coincides with that of the equipotential surface. This provides a natural basis on which the reference clock rate could be defined at Mars's equator, so that all atomic clocks on the areoid would then beat at the same rate. 

We will assume Mars’s center of mass to be the origin of a locally inertial freely falling frame.
This identification is valid provided that the coupling of Mars’s quadrupole and higher-order gravitational moments with the external tidal fields of neighboring bodies—including the Sun, the other planets, and Mars's satellites Phobos and Deimos—is neglected and lies beyond the scope of this study.
Within this approximation, the scalar invariant near Mars is
\be 
-ds^2=-\left(1+\frac{2\Phi_M}{c^2} \right)(cdt)^2+\left(1-\frac{2\Phi_M}{c^2}\right)(dx^2+dy^2+dz^2)\,,
\label{mars_local}
\ee
where the subscript $M$ denotes Mars. $dt$  is the coordinate time interval---analogous to $\mathrm{TCG}$ but defined  for Mars---while $dx$, $dy$, and $dz$ represent the coordinate spatial intervals. A clock at rest on the areoid, at the position $\rvec_M$ relative to the center of mass of Mars, has a velocity 
\be 
\vvec_M=\omegavec_M \times \rvec_M\,,
\ee 
where $\omegavec_M$ is the angular rotational velocity of Mars.  Then, to order $c^{-2}$, the proper time on a clock on the areoid is given by
\be
(d\tau_M)^2=\bigg( 1+\frac{2 \Phi_M(\rvec_M)}{c^2}-\frac{(\omegavec_M \times \rvec_M)^2}{c^2}\bigg)(dt)^2\,,
\ee
where $\Phi_M(\rvec_M)$ is the gravitational potential evaluated at a point $\rvec_M$.  
%We are ignoring the tidal influence of Phobos and Deimos as their effect on clocks near Mars can be shown to be negligible.
Taking a square root and expanding,
\be 
d\tau_M=dt\bigg(1+\frac{\Phi_M(\rvec_M)}{c^2}-\frac{(\omegavec_M \times \rvec_M)^2}{2 c^2}\bigg)\,.
\ee
The relativistic contributions inside the parentheses are approximately constant on the  areoid, as the areoid is assumed to be a surface of hydrostatic equilibrium, then the value of $L_M$ can be evaluated approximately on the equator.  Contributions from higher multipole moments will be very small.
We can define an areocentric constant $L_M$ by
\be 
-L_M=-\frac{GM_M}{a_M c^2}-\frac{1}{2}\frac{GM_M J_{2M}}{a_M c^2}-\frac{\omega_M^2 a_M^2}{2 c^2}\approx -1.4078 \times 10^{-10}= -12.16  \ \mu{\rm s\ day}^{-1}\,,
\label{eq:lmdef}
\ee
where $a_{M}$ is Mars's reference radius used in both models as described below, and $J_{2M} \approx 1.95661 \times 10^{-3}$ is Mars's quadrupole moment coefficient~\cite{Konopliv2016,goossens2025}. 

There are two available models of Mars's gravity field, expanded in spherical harmonics up to degree and order 120. The first model is MRO120D, which is based on Doppler tracking data from the Mars Reconnaissance Orbiter (MRO)~\cite{Konopliv2016}. The second model, known as the Goddard Mars Model (GMM-3), is generated from X-band tracking data acquired by NASA's Deep Space Network stations, using information from three NASA spacecraft: Mars Global Surveyor, Mars Odyssey, and MRO~\cite{goossens2025,goossens2025_data}.
Both models use the same reference radius. The gravitational parameters for Mars are essentially the same in the two models.  The important parameters for the calculation are given in Table~\ref{tab:mars_grav}.
%yielding a value of $L_M = 1.40781 \times 10^{-10}.$
\begin{table}[]
    \centering
    \begin{tabular}{lllc}
    \hline
    \hline
    Model &\hspace{2cm} $GM_M$, gravitational parameter $m^3/s^2$ & Average Potential on Equator $m^2/s^2$ & $L_M$ \\ 
    \hline
    GMM-3  &\hspace{2cm}  $4.282837285418776 \times 10^{13}$ & $-1.2623833543660771  \times 10^{7}$ &$1.40781 \times 10^{-10}$ \\ 
    MRO120D &\hspace{2cm}  $4.282837566396 \times 10^{13}$ &$ -1.261052892342 \times 10^{7} $&$1.40781 \times 10^{-10}$ \\
\hline
\end{tabular}
    \caption{Parameters for two models of Mars gravity potential with reference radius $3.396 \times 10^6$ m.}
    \label{tab:mars_grav}
\end{table}
 Rescaling 
$t \rightarrow t(1-L_M)$, the coordinate time for the Mars-centric reference frame can be made to agree with proper time elapsing on an ideal clock on the areoid.
%We refrain from making any such recommendations, as there are other choices.

\section{Comparison of proper times on Mars and Earth surfaces}
The Sun accounts for more than 99 percent of the mass of the solar system. It is instructive to think about an observer at rest and sufficiently far out from our solar system (``at infinity'') to not be influenced by any external gravitational field. In general relativity, the clock at infinity sets the upper bound on the ticking rate of any clock; no physical clock in a gravitational field can tick faster, apart from comparisons affected by the observer's frame or coordinate choice~\cite{einstein96}.  The clock carried by such an observer would realize the notion of time in Newton's theory.  For a simple model of the rate of a clock on Earth's geoid, we neglect the eccentric motion of the Earth, solar motion, and the influence of solar system bodies other than Earth and the Sun.  Due to the Sun's gravitational potential, this clock would tick faster than an ideal clock at Earth's position by $-GM_s/(1\,\mathrm{AU}\,c^2)$, while due to the motion of Earth around the Sun with velocity such that $V_e^2=GM_s/(1\,{\rm AU} )$, there would be a second-order Doppler shift of $-GM_s/({1\,\rm AU}\,2c^2)$.  Combining these contributions leads to an estimate for the value of the constant $L_C$, which is the fractional rate of Barycentric Coordinate Time (TCB) relative to the rate of TCG:
\be 
-L_C\approx -\frac{3}{2} \frac{GM_s}{1\,{\rm AU}}\frac{1}{c^2}=-1.48059 \times 10^{-9}=-1279.23 \ \mu{\rm s\ day}^{-1}\,,
\ee
where we have used the values of the gravitational parameters from Table~\ref{tab:const}.
Also, a clock on Earth's geoid will experience additional fractional frequency shifts due to Earth's gravitational potential and rotation.  Keeping potential contributions up to quadrupole terms leads to an estimate for the value of $L_G$,  which is the rate of TT relative to TCG \cite{ashby03}
\be 
-L_G \approx -\frac{GM_e}{a_e c^2}-\frac{1}{2}\frac{GM_e J_2}{a_e c^2}-\frac{\omega_e^2 a_e^2}{2 c^2}\approx -6.96927 \times 10^{-10}=-60.21  \ \mu{\rm s\ day}^{-1}\,,
\label{eq:lg_def2}
\ee
where $J_2=1.08263 \times 10^{-3}$ is the Earth's quadrupole moment coefficient, $a_e=6.3781370\times 10^{6}~\rm{m}$ is the Earth's equatorial radius. Combining these contributions gives the rate of the clock at infinity relative to clocks on Earth's geoid:
\be 
L_G+L_C\approx 1339.45 \ \mu{\rm s\ day}^{-1}\,,
\label{eq:lg_to_inf}
\ee
while using the defined values for $L_G$ and $L_C$ (see Table~\ref{tab:const}), gives $1339.64~\mu {\rm s }~{\rm day}^{-1}$.
Proceeding with similar assumptions for the calculation of the rate of a clock at infinity with respect to a clock at the position of Mars, the latter would run slower by an amount determined by a ``Martian'' $L_C$, which we denote by
\be 
-(L_C)_M \approx -\frac{3}{2}\frac{GM_s}{R_M c^2}=- 839.4\ \mu{\rm s\ day}^{-1}\,,
\ee
where we have used the semi-major axis of Mars, $R_M \approx 1.524\,AU$. The rate of a clock at infinity relative to a clock on Mars's areoid, using Equation~(\ref{eq:lmdef}), is then
\be 
L_M+(L_C)_M \approx 851.55 \ \mu{\rm s\ day}^{-1}\,,
\label{eq:lm_to_inf}
\ee
This implies that the rate of a clock on Mars's areoid relative to the rate of a clock on Earth's geoid, using Equation~(\ref{eq:lg_to_inf}) and Equation~(\ref{eq:lm_to_inf}), is 
\be 
L_G+L_C-(L_M+(L_C)_M ) \sim \frac{d\tau_M-d\tau_e}{d\tau_e} \approx 488.08 \ \mu{\rm s\ day}^{-1}\,,
\label{clk_comp1}
\ee
where $d\tau_e$ and $d\tau_M$ are the proper time differences measured by ideal clocks on the geoid and areoid, respectively.
%In the remainder of this section, we shall calculate this rate difference more accurately, taking into account the orbital eccentricities of Earth and Mars.  In the following section, we discuss the effect of solar tides on the motion of Earth and Moon, and their effects on the Mars-Earth clock rate difference.

\begin{table}
    \centering
    \begin{tabular}{ll}
    \hline
    \hline
       $GM_e$, Earth's gravitational parameter\dotfill  & $3.98600435507\times 10^{14}~{\rm m}^3{\rm s}^{-2}$  \\
       $GM_m$, Moon \dotfill\hspace{4cm} & $4.902800118 \times 10^{12}~{\rm m}^3{\rm s}^{-2}$ \\           
       $GM_s$, Sun \dotfill &  $1.32712440041279419\times10^{20}~ {\rm m}^3{\rm s}^{-2}$\\
       $GM_M$,  Mars-system \dotfill & $ 4.2828375816\times10^{13} ~ {\rm m}^3{\rm s}^{-2} $~\cite{Konopliv2016}\\
       $-\Phi_{0M}/c^2, L_M $ \dotfill & $ 1.40781 \times 10^{-10}$, $\sim 12.2~\mu\rm{s}~\rm{day}^{-1}$\, see \S2 \\         
       $-\Phi_0/c^2, L_G $ \dotfill & $ 6.969290134(0) \times 10^{-10}$, \, $\sim 60.2~\mu\rm{s}~\rm{day}^{-1}$~\cite{iau} \\
       $-\Phi_{0m}/c^2, L_m $ \dotfill & $ 3.13881(15)\times 10^{-11} $, \, $\sim 2.71~\mu\rm{s}~\rm{day}^{-1}$~\cite{ashbypatla24}\\
       $L_C$ \dotfill & $1.480813\times 10^{-8}$, \, $\sim 
       1279.43~\mu\rm{s}~\rm{day}^{-1}$~\cite{iau}\\
        $c$, speed of light in vacuum \dotfill & $299792458~{\rm m}{\rm s}^{-1} $ \cite{const1,const2} \\        $e$, assumed eccentricity of the Moon's orbit around the Earth & 0.05490~\cite{moon_fact} \\
        $a$, assumed Earth-Moon semi-major axis distance \dotfill & $ 3.84399 \times 10^8 $~m ~\cite{moon_fact}\\
        AU, astronomical unit \dotfill & $1.495978707\times 10^{11} $~m~\cite{iau2012}\\
       \hline
    \end{tabular}
    \caption{Constants and values. Gravitational parameters list above without references  are estimates from DE440~\cite{park_2021}.}
    \label{tab:const}
\end{table}

But this estimate ignores the gravitational influence of the Moon, the effect of solar tides, and the eccentricity of Mars's orbit. In this section, we provide a fully relativistic formalism that will account for all of the above effects.  
From Equation~(\ref{clk_comp1}), it is easy to verify that clocks on planets at distances greater than $1\,\mathrm{AU}$ and with comparable surface gravity to the Earth will tick faster than clocks on Earth. As an extreme example, and for the more adventurous---if the Sun were a black hole, a clock at the event horizon--- with a radius of $2GM_s/c^2\approx 3~\rm{km}$---would slow down at a rate of $\sim 0.75$ day~$\rm{day}^{-1}$.  This approximation may be compared with the exact spherically symmetric Schwarzschild solution to the Einstein field equations, which yields a rate of -1 (the minus sign indicating a slowdown) when evaluated at the event horizon, where time appears completely frozen to a distant observer.

 It is convenient to work in BCRS to compare the proper times of clocks on the Earth and Mars and use a corresponding coordinate time, TCB, as there is no common center-of-mass as there is for the Earth-Moon system. We shall develop the theory in a form that can be directly compared with data obtained from DE440.  We include relativistic effects only to order $c^{-2}$. Contributions from the tidal potentials of other solar system bodies are not included.
 The results can be evaluated using the positions and velocities of Keplerian orbits for Earth, the Earth-Moon system, and Mars, or the corresponding values from DE440 ephemerides.

The Earth and the Moon orbit around their mutual center of mass. Meanwhile, the center of mass of the Earth-Moon system orbits around the Sun in an approximately Keplerian orbit, and Mars revolves around the Sun in an approximately Keplerian orbit. The notation for velocity is $\Vvec_e$ or $\Vvec_M$ for Earth or Mars, respectively.  Notation for position is $\Rvec_e$, $\Rvec_M$, $\Rvec_m$, $\Rvec_{cm}$, and $\Rvec_s$ for the barycentric positions of Earth, Mars, Moon, the Earth-Moon center of mass, and the Sun, respectively. Position vectors with respect to the Earth and Mars relative to their respective centers of mass are denoted by $\rvec_e$ and $\rvec_M$. The scalar invariant, neglecting the potential or tidal influence of all other solar system bodies, is
\be
-ds^2 = -\left(1+\frac{2\Phi_e}{c^2}+\frac{2 \Phi_m}{c^2}+\frac{2\Phi_M}{c^2}+\frac{2\Phi_s}{c^2}\right)(cdT)^2 
+\left(1-\frac{2\Phi_e}{c^2}-\frac{2 \Phi_m}{c^2}-\frac{2\Phi_M}{c^2}-\frac{2\Phi_s}{c^2}\right)\left(dX^2+dY^2+dZ^2\right),
\ee
 where the uppercase symbols denote quantities expressed in the \textsc{BCRS}, while lowercase symbols are used for quantities in local coordinate systems, such as those centered on the Earth or Mars, see Equation~(\ref{mars_local}). $\Phi_e$, $\Phi_m$, $\Phi_M$, and $\Phi_s$ are the gravitational potentials of the Earth, Moon, Mars, and the Sun.
For a clock with velocity $\Vvec$, expanding to order $c^{-2}$ and taking a square root, the proper time will be
\be\label{propertime}
d\tau=\left(1+\frac{\Phi_e}{c^2}+\frac{\Phi_m}{c^2}+\frac{\Phi_M}{c^2}+\frac{\Phi_s}{c^2}-\frac{V^2}{2 c^2}\right)dT\,,
\ee
where $V=|\Vvec|$, $ds=c d\tau$, and $d\tau$ and $dT$   are the proper and coordinate time intervals respectively. The potentials are evaluated at the position of the clock, and the distances to each contributing body—whose potentials are included—appear in the denominators of the respective terms, as given in Equation~(\ref{pot_term}).

\subsection{Clock on Earth's geoid}

Consider clocks fixed on the surface of the rotating geoid of Earth. 
%The geoid is a surface of approximate hydrostatic equilibrium, so if such clocks are viewed from the local inertial frame, they beat at the same rate, which can be evaluated at the equator. 
The velocity of clocks at the geoid will be 
\be 
\Vvec=\Vvec_e+\omegavec_e \times \rvec_e\,,
\label{vel_add}
\ee
where $\omegavec_e$ is the angular velocity of rotation of Earth, and $\Vvec_e$ is Earth's orbital velocity. Then 
\be 
\frac{\Vvec^2}{2c^2}=\frac{\Vvec_e^2}{2 c^2}+ \frac{\Vvec_e \cdot (\omegavec_e \times \rvec_e)}{c^2}+\frac{(\omegavec_e \times \rvec_e)^2}{2c^2}\,.
\label{vel_term}
\ee
The potential contributions are
\be  
\begin{split}
\frac{\Phi_e}{c^2}+\frac{\Phi_m}{c^2}+\frac{\Phi_M}{c^2}+\frac{\Phi_s}{c^2}
&=-\frac{GM_e}{\vert\rvec_e\vert c^2}\left(1-J_{2}\left(\frac{a_e}{r_e}\right)^2P_2(\cos\theta)\right)\\
&-\frac{GM_m}{c^2\vert \Rvec_e+\rvec_e-\Rvec_m \vert}-\frac{GM_M}{c^2\vert\Rvec_e+\rvec_e-\Rvec_M-\rvec_M\vert}-\frac{GM_s}{c^2\vert\Rvec_e+\rvec_e-\Rvec_s\vert}\,,
\end{split}
\label{pot_term}
\ee
where $P_2(\cos\theta)$ is the Legendre polynomial of degree 2, and $\theta$ is the polar angle measured from the Earth's axis ($\theta=\pi/2$, at the equator).
The last term in Equation~(\ref{vel_term}) and the first term in Equation~(\ref{pot_term}) can be evaluated on Earth's equator, see Equation~(\ref{eq:lg_def1}) and Equation~(\ref{eq:lg_def2}), and combined to be identified with the constant $\Phi_0/c^2$.
%\be 
%-L_G=-\frac{GM_e}{\rvec_e c^2}-\frac{(\omegavec_e \times \rvec_e)^2}{2c^2}=\frac{\Phi_0}{c^2}\,.
%\ee
In the remaining terms in Equation~(\ref{pot_term}), the quantities $\rvec_e$ and $\rvec_M$ are small compared to the other radii.  Expanding the second term on the right-hand side in Equation~(\ref{pot_term}), for example, to first order in $\rvec_e$,
\be 
 -\frac{GM_m}{c^2\vert \Rvec_e+\rvec_e-\Rvec_m \vert}=-\frac{GM_m}{c^2 D}-\frac{GM_m \Dvec \cdot \rvec_e}{c^2 D^3}\,,
 \label{pot_term_1}
\ee
where $\Dvec = \Rvec_m-\Rvec_e$, is the vector distance from Earth to Moon, and $|\Dvec|=D$. The smaller second term in Equation~(\ref{pot_term_1}) will have an approximately 24-hour period and cannot be evaluated unless $\rvec_e$ is specified.  It can be estimated as
\be 
\frac{GM_m r_e}{c^2 D^2} \approx 2 \times 10^{-15}.
\ee
The third term in Equation~(\ref{pot_term}) may similarly be expanded:
\be 
-\frac{GM_M}{c^2\vert\Rvec_e+\rvec_e-\Rvec_M-\rvec_M\vert}=
-\frac{GM_M}{c^2 \vert \Rvec_e-\Rvec_M \vert}+\frac{GM_M (\Rvec_e-\Rvec_M) \cdot (\rvec_e-\rvec_m) }{\vert \Rvec_e-\Rvec_M\vert^3}\,.
\label{pot_term_2}
\ee
The expanded term in Equation~(\ref{pot_term_2}) is less than $10^{-18}$ and can be neglected.  Expanding the last term in Equation~(\ref{pot_term}) in a similar way yields
\be 
-\frac{GM_s}{c^2\vert \Rvec_e+\rvec_e-\Rvec_s\vert} \approx 
-\frac{GM_s}{c^2\vert\Rvec_e-\Rvec_s \vert}-\frac{GM_s(\Rvec_e-\Rvec_s)\cdot \rvec_e}{c^2\vert\Rvec_e-\Rvec_s \vert^3}\,.
\label{pot_term_4}
\ee
The last term has approximately a daily period and is of order $4 \times 10^{-13}$.  Omitting terms that vary with approximate daily periods in Equation~(\ref{vel_term}), the proper time on an earth-based clock is
\ba 
d\tau_e =dT\left(1+\frac{\Phi_0}{c^2}-\frac{GM_s}{c^2 \vert\Rvec_e-\Rvec_s \vert}-\frac{GM_m}{c^2 D}-\frac{GM_M}{c^2 \vert \Rvec_e-\Rvec_M\vert }-\frac{V_e^2}{2 c^2}  \right)\,.
\label{earth_proper_tdb}
\ea

\subsection{Clock on Mars's areoid}
The velocity of a clock at rest on Mars's surface is given by Equation~(\ref{vel_add}) with an appropriate change of notation:
$\Vvec_M$ is the orbital velocity of Mars's center of mass, $\omega_M$ is Mars's angular rotation rate, and $\rvec_M$ is the position of the clock relative to Mars's center of mass.  The contribution due to Mars's gravitational potential, $\Phi_M$, can be evaluated on Mars's equator and combined with the last term in Equation~(\ref{vel_term}) as applied to Mars, see Equation~(\ref{eq:lmdef}).
The contribution from the Moon's potential is
\be 
\frac{\Phi_m}{c^2} = -\frac{GM_m}{c^2\vert \Rvec_M+\rvec_M-\Rvec_m \vert}\approx -\frac{GM_m}{c^2\vert \Rvec_M-\Rvec_m \vert}\,.
\ee
The correction arising from accounting for Mars's radius is less than about $ 3 \times 10^{-18}$ and is negligible.  The contribution from Earth's potential is
\be 
\frac{\Phi_e}{c^2}=-\frac{GM_e}{\vert\Rvec_M+\rvec_M-\Rvec_e  \vert}\approx -\frac{GM_e}{\vert\Rvec_M-\Rvec_e  \vert}\,,
\ee
since the contribution from $\rvec_M$ is similarly negligible.  The contribution from the Sun's potential is
\be 
\frac{\Phi_s}{c^2}=-\frac{GM_s}{c^2 \vert \Rvec_M-\Rvec_s \vert}\,.
\ee
Collecting terms, the proper time of a clock on Mars's surface, analogous to Equation~(\ref{earth_proper_tdb}), is
\ba\label{propertimeMars}
d\tau_M=dT\left(1+\frac{\Phi_{0M}}{c^2}-\frac{GM_s}{c^2\vert \Rvec_M-\Rvec_s \vert}-\frac{GM_e}{\vert\Rvec_M-\Rvec_e  \vert} -\frac{GM_m}{c^2\vert \Rvec_M-\Rvec_m \vert}  -\frac{V_M^2}{2 c^2}\right)\,,
\label{mars_proper_tdb}
\ea
 where, $\Phi_{0M}$ is defined as given in  Equation~(\ref{eq:lmdef}).
The  rate offset between a clock on   Mars's areoid and on Earth's geoid is, therefore, using Equation~(\ref{earth_proper_tdb}) and Equation~(\ref{mars_proper_tdb}) is
\ba\label{fractfreqdiff}
\frac{d\tau_M-d\tau_e}{d\tau_e}=\frac{\Phi_{0M}}{c^2}-\frac{\Phi_0}{c^2}-\frac{GM_s}{c^2\vert \Rvec_M-\Rvec_s \vert}+\frac{GM_s}{c^2 \vert\Rvec_e-\Rvec_s \vert} \hspace{2in} \nonumber\\
-\frac{GM_e}{\vert\Rvec_M-\Rvec_e  \vert} -\frac{GM_m}{c^2\vert \Rvec_M-\Rvec_m \vert}+\frac{GM_m}{c^2 D}+\frac{GM_M}{c^2 \vert \Rvec_e-\Rvec_M\vert } +\frac{V_e^2}{2 c^2}-\frac{V_M^2}{2c^2}.
\ea
 The fractional frequency difference is the measure of how two clocks tick relative to each other. For remote clock comparisons, assuming that within the chosen coordinate chart the coordinate time can be specified without appreciable uncertainty, the clock rate offset given by Equation~(\ref{fractfreqdiff}) is the same as the fractional frequency difference~\cite{null_test18}.
Each of the terms in Equation~(\ref{fractfreqdiff}) can be evaluated using either DE440, or Keplerian models for the orbits of Mars, the Earth-Moon center of mass, and the orbit of the Moon relative to the Earth. Keplerian orbits lie  close to the ecliptic plane, but  DE440 uses the  International Celestial Reference Frame (ICRF), which is aligned with Earth's mean equator and equinox of J2000.0 (direction of the vernal equinox on January 1, 2000, at 12:00:00 TT), and not the ecliptic. The equinox of J2000.0 defines the zero point of right ascension in the ICRF~\cite{kaplan06,urban2013}.  DE440 positions and velocity components are to be rotated by the obliquity of the ecliptic at J2000.0,  $23.4392911^\circ$, to  compare with Keplerian components.   To obtain all the parameters needed for a complete Keplerian model of Mars’s orbit, we matched the instant of perihelion passage with DE440 but otherwise used orbital elements, as listed in Table~\ref{tab:kepel}. The time scale and spatial coordinates used in our computations are those of the DE440 ephemerides.
In the model represented by Equation~(\ref{fractfreqdiff}), even though quadrupole moment coefficients are included for Earth, Moon, and Mars for the comparison of clock rates, gravitational torques on the moments are neglected.

\begin{figure}
    \centering
    \includegraphics[width=\textwidth]{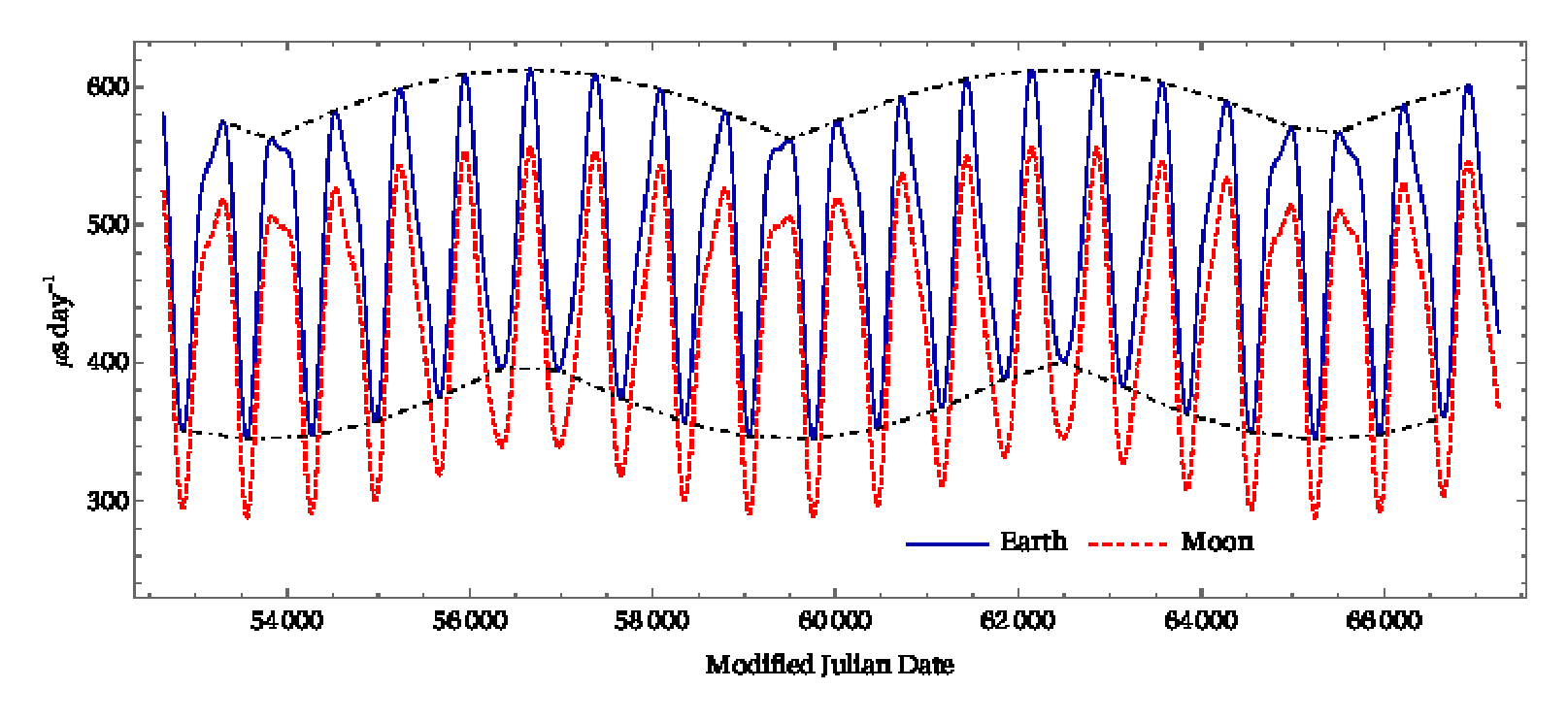}
    \caption{Plots of the  clock rate offsets between a clock on Mars compared to clocks on the Earth and the Moon for $\sim 40$ years starting from MJD~52275 (January 1, 2003), using the DE440.
    %Since the rate offset is computed using proper time differentials {\bf, that are independent of coordinate systems used,} it is convenient to compute t
    The Earth-Moon rate offsets are computed using expressions derived in~\cite{ashbypatla24}. 
    Mars’s $\sim15.8$~yr seasonal opposition cycle consists of three or four aphelic oppositions and three consecutive perihelic oppositions.  These so-called seven Martian synodic periods are not exact,  see discussion at the end of \S3. But every ~79~yr, the pattern of Mars–Earth opposition geometry and seasons recur almost exactly~\cite{mars_opp}.  In order to compute the mean rate offset between Mars and Earth, we use only one $\sim15.8$~yr cycle to estimate the mean rate to be $477.60~\mu\rm{s}~\rm{day}^{-1}$ with a mean oscillation amplitude of $226.79~\mu\rm{s}~\rm{day}^{-1}$ over a Martian period (1.88~yr). The rate offset of Mars with respect to the Moon is $421.55~\mu\rm{s}~\rm{day}^{-1}$ with the same mean oscillation amplitude. The difference between these two rates  is $\approx56.05~\mu\rm{s}~\rm{day}^{-1}$.  Additionally, the amplitude envelope of the clock-rate offset between Earth and Mars, as well as between Moon and Mars, shows a modulation of about $\sim40~\mu\rm{s}~\rm{day}^{-1}$ over a seven Martian synodic cycle.
    } 
    \label{fig:mars1}
\end{figure}

The clock rate offsets computed using DE440 together with the constants listed in Table~\ref{tab:const} are shown in Figure~\ref{fig:mars1}.
Mars's distance from Earth varies between $2.67\,\mathrm{AU}$ and $0.37\,\mathrm{AU}$. The synodic period, which is the time it takes for Mars to return to the same position relative to the Sun as observed from Earth, is \( (T_e^{-1} - T_M^{-1})^{-1} \approx 2.13 \)~yr; \( T_e \) and \( T_M \) represent the orbital periods of Earth and Mars, which are 1~yr and 1.88~yr, respectively.  Because the orbital periods of Mars and Earth are incommensurable, the seven Martian synodic cycles, see Figure~\ref{fig:mars1}, do not repeat exactly~\cite{mars_opp}. 
When Mars and Earth are in opposition and Mars is at perihelion---its closest approach to the Sun--- the Earth-Mars distance reaches a minimum ($\sim0.37\,\mathrm{AU}$). This configuration is known as a perihelic opposition. Conversely, during aphelic opposition, Mars lies near aphelion, and the Earth–Mars distance is at its maximum for an opposition ($\sim0.68\,\mathrm{AU}$). In all oppositions, Earth lies between Mars and the Sun, so that from Earth’s perspective, Mars and the Sun appear in opposite directions. The variability of clock-rate offsets between Earth and Mars over seven synodic cycles is governed primarily by Mars’s orbital eccentricity and the influence of solar perturbation on both planets during these oppositions. From Equation~(\ref{fractfreqdiff}), using DE440 ephemerides, we estimate that the clocks on Mars tick faster by $477.60~\mu\rm{s}~\rm{day}^{-1}$ with a mean additional amplitude variation of $226.80~\mu\rm{s}~\rm{day}^{-1}$ over a Martian period compared to clocks on Earth, and by $421.55~\mu\rm{s}~\rm{day}^{-1}$ and with same amplitude compared to clocks on the lunar surface. Additionally, there is modulation of $\sim 40~\mu\rm{s}~\rm{day}^{-1}$ over a seven-synodic period of $\sim15.8$~yr. 
%{\bf In estimating the rate offsets of Equation~(\ref{fractfreqdiff}),the residuals --- obtained by subtracting the  Keplerian model from DE440 ephemerides --- remain within a few hundred nanoseconds, and the main period of the residuals is that of the Moon's orbit around the Earth.}

The residuals obtained by differencing the results of evaluation of  Equation~(\ref{fractfreqdiff}) using the Keplerian model and DE440 show a modulation corresponding to the period of the Moon around the Earth, with varying amplitude of a few hundred ns~day$^{-1}$. This large effect is due to the inadequacy of the simple Keplerian models. It arises from the neglect of solar tides on the Earth and Moon. That is, solar tides acting on the Earth-Moon system cause the position and velocity of Earth to change with the orbital period of rotation of the Earth-Moon line. The modified position and velocity of the Earth then give rise to a large effect on the Earth and Mars clock comparison that reduces the residual. This is quite a departure from the Earth-Moon rate offset that was computed in ~\cite{ashbypatla24}, where the tidal concerns were less severe.

In the following section, we introduce a formalism   that incorporates the effects of solar tides on the Earth–Moon system and demonstrate, using a numerical integration scheme, that this approach significantly improves the accuracy of rate calculations for both Earth–Moon and Earth–Mars comparisons.

\section{Solar Tidal Effects on the Earth-Moon System}
\begin{figure}[H]
    \centering
    \includegraphics[width=\textwidth]{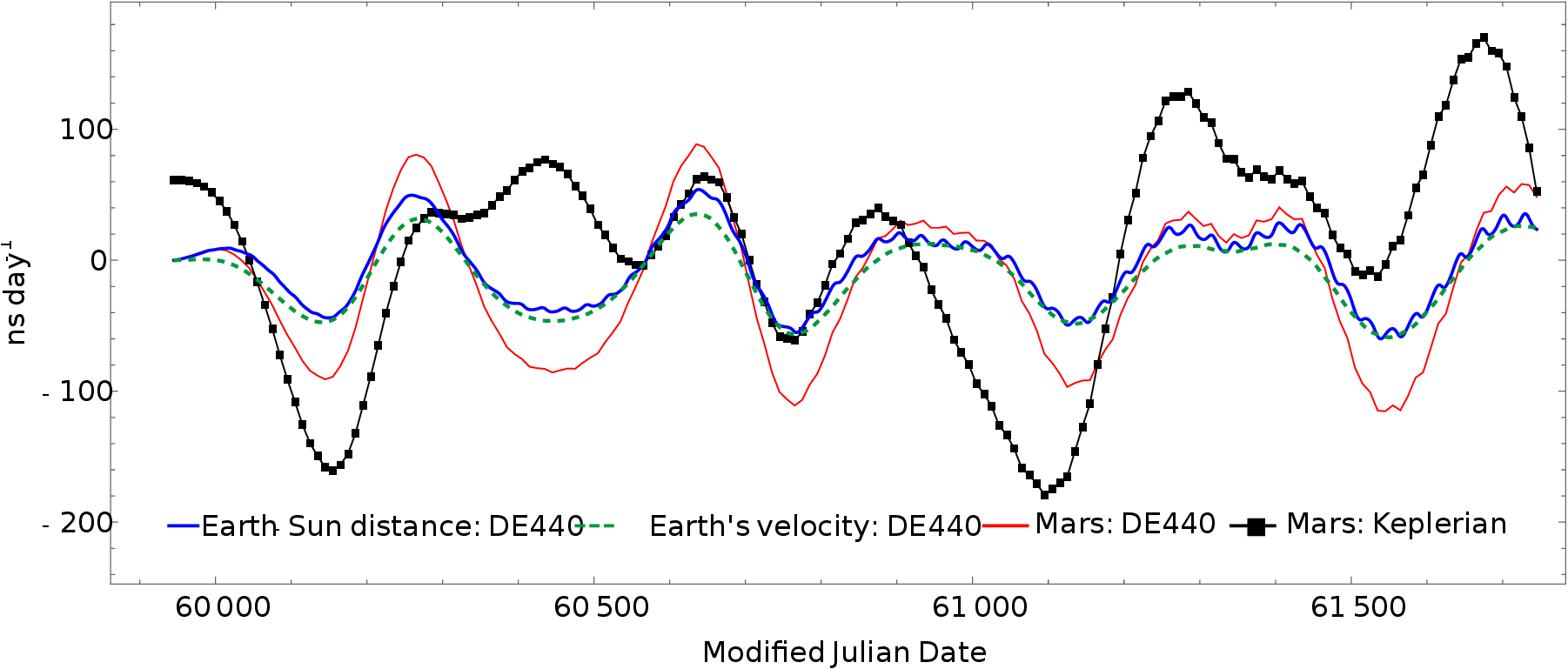}
    \caption{ The residuals from the evaluation of Equation~(\ref{fractfreqdiff})—which represents the rate offset between Mars and Earth—are obtained by differencing the solar-tide–corrected models from the DE440 solution. The Keplerian model (line on solid squares) is limited in that it does not account for the effects of solar tides on Mars (mostly during oppositions), although the Earth and Moon orbits are corrected for such effects. The residuals indicate that the maximum variation in the rate offsets can reach up to $\pm 175~\rm{ns}~\rm{day}^{-1}$ over the course of a few Martian years. Using DE440 ephemerides for Mars (solid line with filled circles) constrains the residuals to within $\pm 100~\rm{ns}~\rm{day}^{-1}$. The solid and dashed curves illustrate the effect of using DE440 models for the Earth's potential and Doppler terms in Equation~(\ref{fractfreqdiff}). These results show that the Doppler contribution, after applying solar-tide corrections to the Earth's motion, amounts to $\pm 45~\rm{ns}~\rm{day}^{-1}$, while the error associated with the Earth-Sun distance estimation is $\pm 55~\rm{ns}~\rm{day}^{-1}$. Thus, the neglected motion of the Sun in the heliocentric coordinate system contributes to errors in the Earth–Sun distance and, indirectly, in the estimation of Earth’s velocity. The clock rate offset residuals stay within a few nanoseconds over $\sim100$ days when the velocity and position are updated using solar tide-corrected models together with the DE440 ephemerides for Mars. } 
\label{fig:mars2}
\end{figure}

\begin{figure}[H]
  \centering\begin{subfigure}{0.93\textwidth}
    \includegraphics[width=0.93\linewidth]{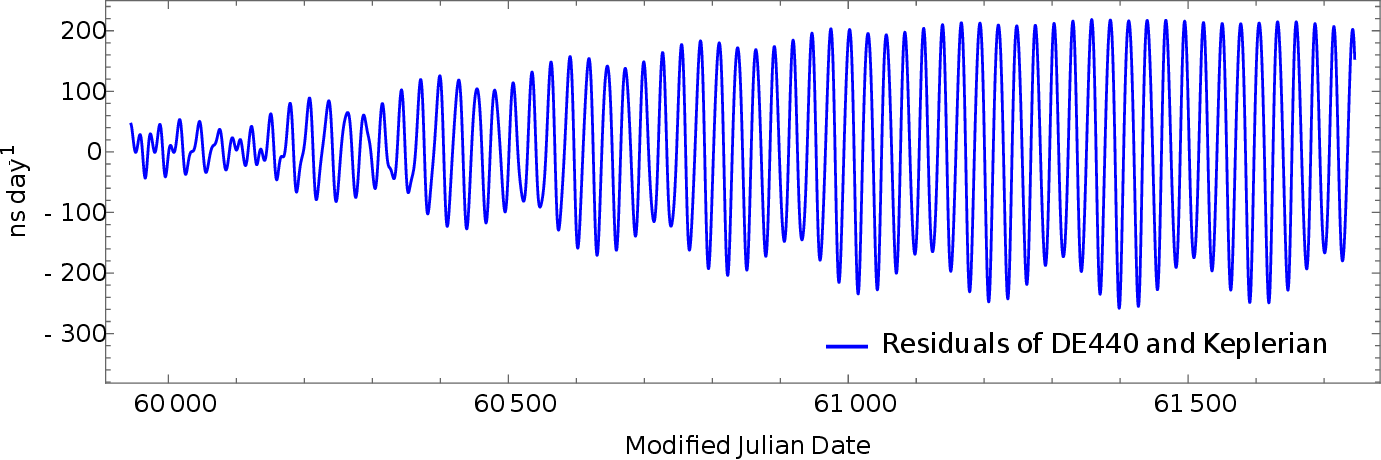}
   \caption{ Earth-Moon rate offsets from differencing Keplerian Models from DE440: residuals of Earth–Moon clock-rate offsets in the center-of-mass frame, diverging to $\pm 50~\rm{ns}\,\rm{day}^{-1}$ and $\pm 200~\rm{ns}\,\rm{day}^{-1}$ over timescales ranging from days to years, with oscillations at the lunar periods.}
  \end{subfigure}\\[1ex]  
  \begin{subfigure}{0.92\textwidth}
    \includegraphics[width=0.92\linewidth]{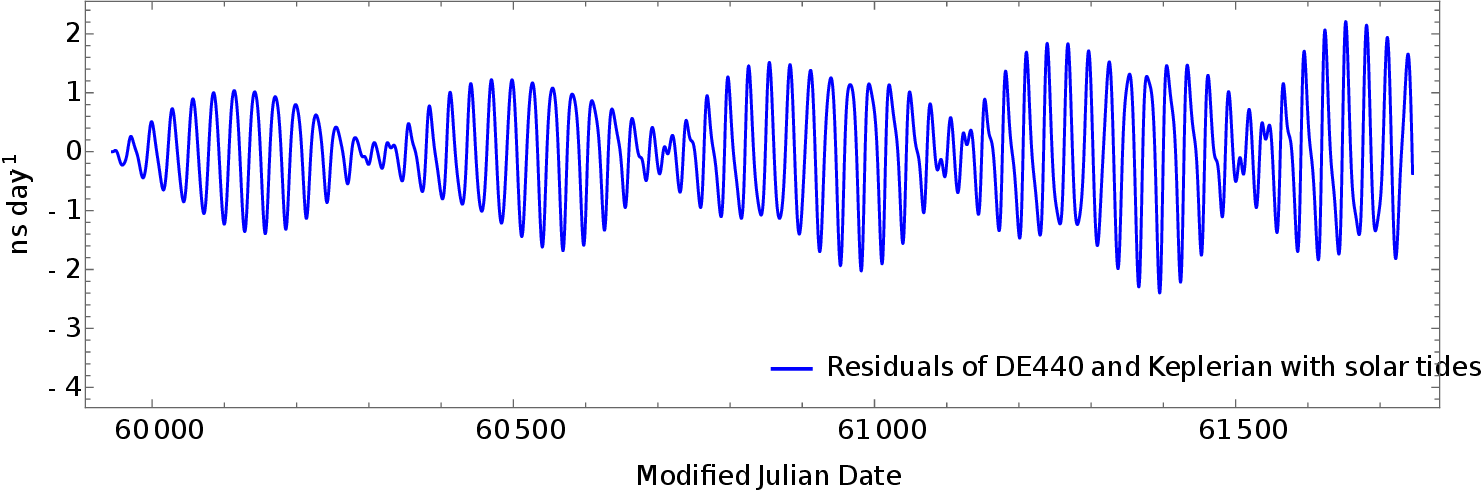}
\caption{ Earth-Moon rate offsets from differencing solar-tide corrected Keplerian models from DE440: residuals remain within $\pm 3~\rm{ns}\,\rm{day}^{-1}$ over timescales of several years, with oscillations at the lunar periods. }
  \end{subfigure}\\[1ex]
  \begin{subfigure}{0.89\textwidth}
    \includegraphics[width=0.89\linewidth]{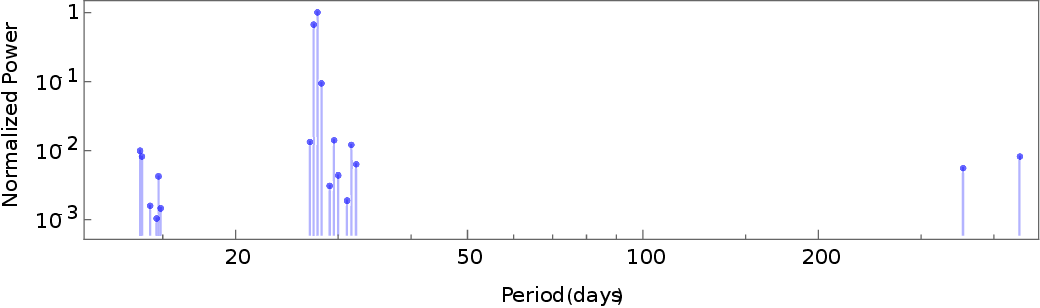}
   \caption{ The Fourier transform of the solar-tide–corrected residuals reveals spectral components at half and full synodic ($\approx 29.5$\,days) and sidereal ($\approx 27.3$\,days) periods of the Moon, together with annual modulations.}
  \end{subfigure}
 \caption{Earth–Moon clock-rate offset residuals from Keplerian models with and without solar-tide corrections.}
\label{fig:mars3}  
\end{figure}

\begin{figure}[H]
    \centering
    \includegraphics[width=0.70\linewidth]{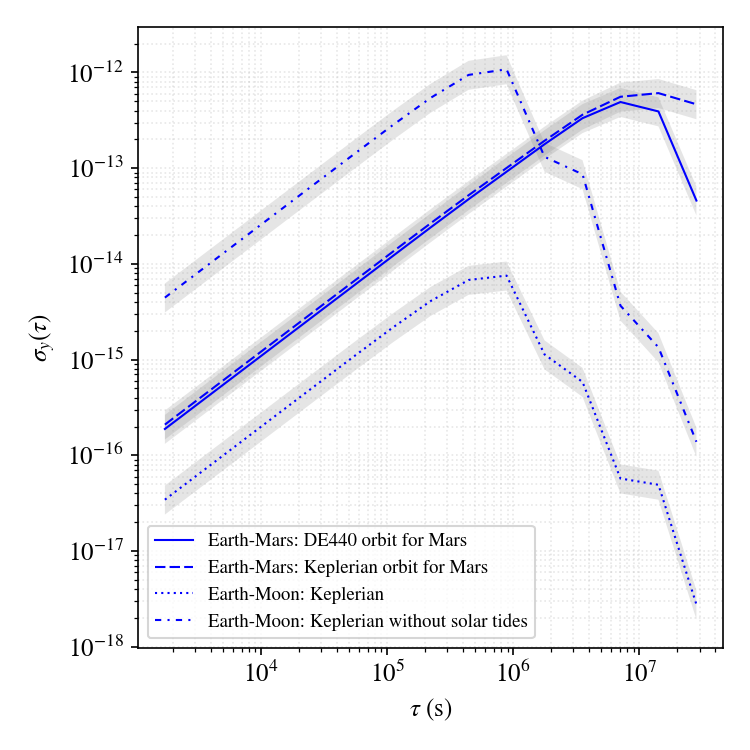}
    \caption{Modified Allan deviation of fractional frequency residuals~(with respect to DE440) as a function of averaging time for the estimated rate offsets for  the Earth and Mars and Earth and Moon comparisons. All fractional frequency data presented here were obtained using our solar tides model in $\S4$, except for the one model specifically identified in the plot legend.  The positive slope of  (+1) corresponds to frequency drifts. The negative slopes mostly fall in the regime of white phase modulation (WPM) noise with slope, $\sim -3/2$, that may be attributed to noise in the numerical integration and orbit estimation. Negative slopes are also indicative of phase/frequency modulation with periods comparable to the averaging interval.
    The Earth–Moon rate offset improves by nearly two orders of magnitude for both short-term (a few days) and long-term (several tens to 100 days) averaging once solar-tide corrections are applied. Numerical integration noise is found to be negligible and within confidence intervals set to $3\sigma$  (shading in plots).  Using only the frequency drift regimes, the Earth and Mars clock comparisons can be constrained to within $\sim 5\times 10^{-13}$ after averaging over 10~days and to within $\sim 10^{-13}\,(\pm 100\,\rm {ns}~\rm{day}^{-1})$ after averaging over 100 days.}
    \label{fig:adev}
\end{figure}

The Sun's influence on the Earth-Moon system manifests not only in their orbital dynamics  (periodic, reversible oscillations in orbital elements) but also through subtle tidal perturbations that affect their long-term evolution  (irreversible secular changes). The physical origin of solar tides lies in the differential gravitational pull exerted by the Sun on the Earth-Moon system.  Because the Earth and Moon are separated by a finite distance, the tidal acceleration due to the Sun will perturb the mutual motion of the Earth and the Moon. This manifests as a quadrupole-like tidal field that modifies the effective potential governing the Moon's motion around Earth. In order to simplify the problem, we shall treat the Earth and Moon as point masses moving around the Sun.  The relative motion of the Earth–Moon system with respect to the Sun is described in a heliocentric coordinate system. The Lagrangian that captures the mutual gravitational interactions among the Earth, Moon, and Sun can be written as 
\ba 
L=\frac{1}{2}M_e\left(\frac{d\Rvec_e}{dt} \right)^2+\frac{1}{2}M_m\left( \frac{d \Rvec_m}{dt}\right)^2
+\frac{G M_e M_m}{D}+\frac{GM_s M_e}{\vert \Rvec_e \vert }+\frac{GM_s M_m}{\vert \Rvec_m \vert }\,,
\label{eq:em_lag1}
\ea
 where the Sun is assumed to be stationary, and with no corresponding kinetic energy term in Equation~(\ref{eq:em_lag1}).
Transforming coordinates to center-of-mass $\Rvec_{cm}$ and relative $\Dvec$ coordinates,
\be 
\Rvec_{cm}=\frac{M_e \Rvec_e +M_m \Rvec_m}{M_T} \quad  \rm{and}  \quad \Dvec=\Rvec_m-\Rvec_e \,,
\ee
where $M_T=M_e+M_m$, and expanding the solar contributions about the center of mass, the Lagrangian becomes
\ba
L=\frac{1}{2}M_T\left(\frac{d \Rvec_{cm}}{dt}\right)^2+
\frac{M_e M_m}{2M_T}\left(\frac {d \Dvec}{dt} \right)^2+\frac{G M_e M_m}{|\Dvec|}+\frac{G M_s M_T}{|\Rvec_{cm}|} \nonumber \hspace{2in}\\
+\frac{GM_s M_e M_m}{2M_T}
\left(\frac{3 (\Rvec_{cm} \cdot \Dvec)^2-|\Rvec_{cm}|^2 |\Dvec|^2}{\vert \Rvec_{cm} \vert^{5} }\right)\,.
\ea
From this Lagrangian, which does not depend explicitly on time, the equations of motion of the center of mass point and the Earth-to-Moon distance are easily derived using standard Lagrangian dynamics, giving twelve coupled equations  for positions and velocities of Earth and Moon.
Assigning a component notation for the vectors $\Dvec\equiv D^i $ and $\Rvec_{cm}\equiv R_{cm}^i$, where $i=1,2,3$ denote the rectangular Cartesian components of position, velocity, and acceleration, we obtain

\ba
\ddot{R}_{cm}^i =-\frac{GM_s R_{cm}^i}{|\Rvec_{cm}|^3} + \frac{GM_s GM_e GM_m (3D^i \Rvec_{cm} \cdot \Dvec - R_{cm}^i \Dvec \cdot \Dvec)}{ (GM_T)^2 |\Rvec_{cm}|^5} \hspace{2.5in} \nonumber  \\
   -\frac{5 GM_s GM_e GM_m R_{cm}^i \left( 3 (\Rvec_{cm} \cdot \Dvec)^2 - (\Dvec \cdot \Dvec)  (\Rvec_{cm}\cdot\Rvec_{cm}) \right)}{2(GM_T)^2 |\Rvec_{cm}|^7 },
\label{eq:tide_r}
\ea

\be
\ddot{D^i} = -\frac{GM_T D^i}{|\Dvec|^3} + \frac{GM_s ( 3 R_{cm}^i\Rvec_{cm} \cdot \Dvec -  D^i \Rvec_{cm}\cdot\Rvec_{cm})}{|\Rvec_{cm}|^5 }.
\label{eq:tide_d}
\ee
The acceleration components can be numerically integrated to estimate the  position and velocity components using a symplectic integration algorithm such as the Störmer-Verlet method~\cite{hairer2003}. Using this formalism, we can easily derive the distances in component forms for terms that do not involve Mars in Equation~(\ref{fractfreqdiff}). The coordinates in the Earth-Moon center of mass frame and the barycentric frame can be related using the following expressions.
\ba
\Rvec_e - \Rvec_s \equiv R_{cm}^i-\left(\frac{GM_m}{GM_T}\right)D^i\,,  \quad 
\Rvec_m - \Rvec_s \equiv R_{cm}^i+\left(\frac{GM_e}{GM_T}\right)D^i.
\label{eq:comp1}
\ea

\ba
\Rvec_M - \Rvec_m \equiv \Rvec_M - \Rvec_s -\left(  R_{cm}^i+\left(\frac{GM_e}{GM_T}\right)D^i\right)\,,
\quad
\Rvec_M - \Rvec_e \equiv \Rvec_M - \Rvec_s -\left(  R_{cm}^i-\left(\frac{GM_m}{GM_T}\right)D^i\right).
\label{eq:comp2}
\ea
Similar relationships can be derived for the velocity components. The Keplerian  orbit for Mars was constructed using the orbital elements given in Table~\ref{tab:kepel} 
\begin{table}
    \centering
    \begin{tabular}{ll}
    \hline
    \hline
       semi-major axis \dotfill &\hspace{7cm}  $1.52366231\,\mathrm{AU}$                \\
       eccentricity \dotfill  &\hspace{7cm}  $0.09341233$ \\
       inclination (with ecliptic) \dotfill &\hspace{7cm} $1.85061^{\circ} $        \\
       longitude of the ascending node, $\Omega$ \dotfill &\hspace{7cm} $49.57854^{\circ}$    \\
       argument of periapsis, $\omega$ \dotfill &\hspace{7cm} $286.4623^\circ$  \\
       time of perigee passage \dotfill &\hspace{7cm} MJD~60438.44733      \\  
       \hline
    \end{tabular}
    \caption{Orbital elements used for generating a Keplerian orbit for Mars. The time of perigee passage is obtained by matching the perigees of Keplerian orbit for Mars with DE440 ephemerides. All other orbital parameters are obtained from~\cite{NASA_NSSDC_MarsFactSheet}.}
    \label{tab:kepel}
\end{table}
for only the Martian orbit. The orbits for the Earth and Moon are constructed by numerically integrating Equation~(\ref{eq:tide_r}) and  Equation~(\ref{eq:tide_d}) and substituting in Equation~(\ref{eq:comp1}) and Equation~(\ref{eq:comp2}). The starting values of the numerical integration can be set using Keplerian or DE440  position and velocity components.
%The Keplerian orbits are constructed by matching the perigee of the orbits that are obtained from DE440. 
 Figure~\ref{fig:mars2} shows the residuals of the Earth–Mars clock-rate offsets, obtained by subtracting the Keplerian evaluations of Equation~(\ref{fractfreqdiff})---with and without solar-tide corrections---from the DE440 evaluation.
The improvement in the estimates of clock rates for the Earth-Moon system as discussed in \cite{ashbypatla24} is given in Figure~\ref{fig:mars3}.
Since we are comparing clock rates, two-sample variances that are typically used to characterize the stability of oscillators may be used to analyze the spectral components present in the clock rates given by Equation~(\ref{fractfreqdiff}), as they are the fractional frequency differences as well, and are plotted in Figure~\ref{fig:adev}~\cite{riley_allan,wallin_allantools_2024}.
The modified Allan deviation (MDEV) extends the standard Allan deviation by incorporating phase averaging, which enhances its sensitivity to phase noise processes. In particular, it can distinguish between white phase modulation (PM) and flicker PM noise, both of which are critical in short-term remote clock comparisons. This capability makes MDEV especially valuable for characterizing the stability of time-transfer links used for comparing remote clocks.

 The modified Allan deviation (MDEV), although traditionally used to characterize noise processes in oscillators, is employed here primarily to quantify drift rates and to reveal possible signatures of phase and frequency modulation over the relevant averaging intervals. The relevance of two-point variances such as MDEV will grow as clocks and time-transfer links are deployed on the Moon and Mars, where benchmarking these links against theoretical predictions becomes essential.  

\pagebreak

\section{Conclusions}
%\hypertarget{sect11}{}

In this paper, we build upon the framework that was presented in \cite{ashbypatla24} and apply it to the planet Mars by including the effects of the solar tides on the Earth-Moon system. We show that the Earth and Mars clock comparison is more complex than the Earth-Moon system. The large distance separation and the variation of this distance between Earth and Mars, over a synodic period of 2.13~yr, from a maximum of $\sim 2.7\,\mathrm{AU}$ and a minimum of $\sim 0.4\,\mathrm{AU}$ result in correspondingly large modulation in estimated clock rates of $40~\mu\rm {s}~\rm{day}^{-1} $ on top of the $226~\mu\rm {s}~\rm{day}^{-1} $ amplitude over the duration of Mars's period.

Extending the calculation using Equation~(\ref{fractfreqdiff})  for comparing rate offsets of Mars and Moon, we find that on average, the clocks on Mars tick faster compared to clocks on the Moon by $421.5~\mu\rm {s}~\rm{day}^{-1} $. By numerically estimating the tidal perturbation, the estimates for the rate offset for the Earth-Moon system were improved by almost two orders of magnitude compared to those presented in \cite{ashbypatla24}.  Even after incorporating the solar-tide model for the Earth–Moon system, the rate offsets for Earth–Mars and Moon–Mars comparisons retain an inaccuracy of $\pm100~\rm {ns}~\rm{day}^{-1}$ over long timescales, owing to unmodeled planetary perturbations and the neglected motion of the Sun relative to the DE440 ephemerides. Further reduction of these residuals requires either accounting explicitly for solar motion in Equation~(\ref{eq:em_lag1}) or updating the Earth's position and velocity components at intervals of order $\sim 100$\,days. In practice, this would involve supplying accurate position and velocity values at those intervals, restarting the numerical integration from each update point, and applying numerical smoothing techniques to ensure continuity across the boundaries between successive segments~\cite{gelb74,press2007}. By contrast, for the Earth–Moon rate comparison, the residuals remain within a few nanoseconds even without frequent updates of position and velocity.   

%Together with the frequency drift indicated by a slope of +1 in the graphs of Figure~\ref{fig:adev}, this signifies that further reduction in the residuals depends on improvement in the model of the Moon’s motion.}     

We note that our current model does not incorporate several relativistic effects that could become relevant in more precise analyses.  These include orbital precessions, interactions involving gravitational multipole moments (such as the quadrupole fields of Earth and Mars), Lorentz-type velocity-dependent corrections, and relativistic effects associated with length contraction and their indirect influence on proper time (see, e.g.,~\cite{Kopeikin1988,iau03, urban2013,kopeikin2020}). While these effects are expected to be smaller than the dominant terms considered here, their cumulative impact may become significant in future high-precision timing models and interplanetary navigation. Assessing the full extent of their influence, particularly in the context of establishing a robust and scalable time standard for Mars, will require further investigation.

\begin{acknowledgments}
B.R.P. acknowledges funding from the NASA grant NNH12AT81I.  We thank the anonymous referee for providing valuable suggestions that have helped improve this paper.
%We are grateful to Elizabeth Donley, who carefully and critically reviewed the manuscript and provided valuable suggestions. We would also like to express our gratitude to Cheryl Gramling for initiating discussions on lunar time. We extend our sincere thanks to Roger Brown, Thomas Heavner, Judah Levine, Jeffrey Sherman, and Daniel Slichter for their review of the manuscript. \textbf{We thank the anonymous referee for providing valuable suggestions that have helped improve this paper}. 
This work is a contribution of NIST and is not subject to US copyright.
\end{acknowledgments}

\bibliography{mars_2025} 
\bibliographystyle{aasjournal}    % Produces the bibliography via BibTeX.

\end{document}